# Inverse Design of All-dielectric Parallel-Plane Mirror Resonator

M.Tutgun[1,*], D.Yilmaz[1], A.Yeltik[2], M. Turduev[3] and H. Kurt[1]

[1]Department of Electrical and Electronics Engineering, TOBB University of Economics and Technology, Ankara 06560, Turkey
[2]Department of Physics, University of Cambridge, Cambridge CB3 0HE, UK
[3]Department of Electrical and Electronics Engineering, TED University, Ankara 06420, Turkey
*e-mail: mtutgun@etu.edu.tr

ABSTRACT

In this study, we used parallel plane mirror resonator analogy to design all-dielectric optical resonator that is performed with the objective-first inverse design algorithm. Confinement of the light is succeeded via predefined objective function to create symmetric mirror regions for the fundamental transverse-electric polarization mode. The algorithm creates bandgap in random structure to obtain resonance peak at the desired wavelength without any intuitive scanning of the parameter space. The first order mode is confined on the cavity region at pre-defined wavelength. The obtained structure is analysed by using the two-dimensional finite-difference time-domain method. The proposed structure has a compact configuration with a footprint of 13.394 x 0.592 μm$^2$.

I. INTRODUCTION

All-dielectric photonic structures are promising candidates for a wide range of photonic and optoelectronic applications such as nonlinear optics, optical filtering, chemical/bio-sensing, opto-mechanics, and quantum information processing [1-3]. The optical resonators which allow the control of the radiation to be compressed into certain modes, have an important place for the photonic integrated structures due to its high performance in light matter interaction [4,5]. Optical resonators are used in many applications including micro/nano-scale sensors, lasers, oscillators, and photodetectors [6-9]. Most of the photonic resonator designs in the literature have been designed with the use of bandgap manipulation by creating a defect in the structure of the photonic crystal and using layered structures to create distributed Bragg reflection [10-12]. Many of these designs are performed by obtaining resonance modes in the band gap and intuitively performed by testing structural parameters. In addition, optical resonator specifications of disordered structures are studied due to their unique ability of storing photon energy [13,14]. Also, there are studies to manipulate transverse modes in optical resonators and investigate resonant-cavity modes [15-17].

Recently, various methods of inverse design are extensively studied in photonics since they have great potential to provide plenty of degree of freedoms in structural optimization using large numbers of parameters that is not possible with conventional approaches [18,19]. With this motivation, an effective design approach named as objective-first (OF) inverse design algorithm has been recently proposed along with the implementation of various integrated photonic components exhibiting high performances [20]. Differently from the previously reported approaches, the OF algorithm uses a significantly broad parameter space, and provide higher functionality, design flexibility, effectiveness in terms of computational cost and comparably small footprint size of the device. Furthermore, it is a highly promising technique for manufacturable photonic devices enabled by standard lithography and semiconductor fabrication processes [20, 21]. Therefore, in this study, we propose and demonstrate dielectric mirror-based resonator obtained via the approach of the OF inverse design. The paper is organized as follows. Section II explains the targeted design problem and outlines the optimization method. Then Section III shares the numerical results of optical resonators. Finally, Conclusions are given in Section IV.

## II. OBJECTIVE-FIRST DESIGN APPROACH AND ITS PRINCIPLE OF OPERATION

In general, optimization-based approaches of photonic devices can be considered as forward design problems where intuitively starting from an initial structure, the design is continued by trial and error until the desired result is achieved. Here, by iteratively optimizing the structural parameters and analysing light characteristics (such as transmission/reflection coefficients) one can obtain the possible best result in the defined search domain. On the other hand, as seen in Fig. 1(a), the OF design approach of the photonic device is considered as an inverse problem. The target parameters of the photonic design are desired electromagnetic (EM) field response $(\vec{E}, \vec{H})$ in the OF design technique. Then the dielectric distribution $(\varepsilon)$ of the structure is iteratively obtained. The wave equation, which is given in Eq.1, presents a non-convex problem where both the $\varepsilon$ and $\vec{H}$ are the parametric variables:

$$\nabla \times \varepsilon^{-1} \nabla \times \vec{H} - \mu_0 \omega^2 \vec{H} = \nabla \times \varepsilon^{-1} \vec{J}, \tag{1}$$

Therefore, the OF algorithm is considered as a bi-convex problem using the linearity property of the Maxwell equation in magnetic field $(\vec{H})$ and dielectric function $(\varepsilon)$ separately. In this manner, the steady-state wave equation is divided into two sub-problems, which are field sub-problem and structure sub-problem as shown in Eqs. 2 and 3, respectively:

$$\min_x \left\| A(p)\vec{x} - b(p) \right\|^2 \text{ subject to } f(\vec{x}) = f_{ideal}, \tag{2}$$

$$\min_p \left\| B(\vec{x})p - d(\vec{x}) \right\|^2 \text{ subject to } p_{\min} \leq p \leq p_{\max}, \tag{3}$$

where $A(p) = \nabla \times \varepsilon^{-1} \nabla \times$, $b(p) = \nabla \times \varepsilon^{-1} \vec{J}$ and $p = \varepsilon^{-1}$.

In the OF design approach, the nature of the forward design is manipulated by changing the roles of the target and constraint functions as described in the inverse design method [23]. This eliminates the need for strict adherence of the structure to the wave equation and permits the inequality of $A(p)\vec{x} - b \neq 0$ as can be seen in Eq. 2. The parameter named "*physical residual*" is defined to indicate how the structure is far away from the wave equation. In the OF strategy, our constraint function is defined to keep the physical residual as low as possible. Here, the function $f(\vec{x})$ is the objective function that is defined for the desired device represented in Eq. 2. Also, we have another constraint, the limit of the dielectric function, defined in Eq. 3. Instead of solving the wave equation directly, the targeted photonic design is obtained by alternating between the structure and field sub-problems. At the field sub-problem, we solve $p$ for given $x$ and vice versa (solving $x$ for given $p$) at the structure sub-problem. In other words, OF approach continually alternates between two sub-problems until it reaches some pre-defined stopping point.

The design area consists of individual square shaped cells/pixels and OF approach iteratively determines the value of dielectric constant for each cell. In the parallel reflector-based resonator design, the dielectric limits are predefined as $\varepsilon_{air} \leq \varepsilon \leq \varepsilon_{Si}$ where $\varepsilon$ changes between the value dielectric value of air and Si materials continuously. It is important to note that after determining the desired photonic design having continuous dielectric distribution, a discretization step should be applied to obtain a feasible binary structure composed of only Si and air holes. Here, by using the "binarization cost" method [24], the formation of a discretized structure is achieved.

The optical resonator that intended to design in the proposed study can be considered as the optical counterpart of an electronic resonant circuit that confines and stores light at predefined resonance frequencies [25]. In resonator designs including Fabry-Perot resonators and optical filters, the dielectric mirrors play significant role because they enhance the confinement of light due to efficiently reflecting of the incident light [26, 27]. The schematic illustration of the parallel reflector-based resonator is shown in Fig. 1(b). Here, the standing wave is generated by revealing constructive interference between two parallel plane reflectors. The important structural parameter is the distance between reflectors, *L*, i.e. cavity length, can be defined by Eq. 4[25]:

$$L = m \frac{\lambda_m}{2}, \tag{4}$$

where $m$ is the number of modes and $\lambda_m$ is the wavelength of the resonance modes. The wavelength can be also expressed by the ratio of the speed of light in the cavity medium ($c = \frac{c_0}{n_{eff}}$, $c_0$ is the speed of light in the vacuum and $n_{eff}$ is the effective refractive index of the cavity region) to the resonance frequency ($f_m$) thereby Eq. 5 can be obtained from Eq. 4. The frequency values for resonance modes that are supported by resonator can be driven from Eq. 5 as follows [25]:

$$f_m = m \frac{c_0}{n_{eff} 2L}, \tag{5}$$

It should be noted that if the refractive index is constant inside the resonator, the frequency difference between adjacent modes are constant which is called the free spectral range, but in random structures mode spacing differ because of the refractive index modulation along the structure.

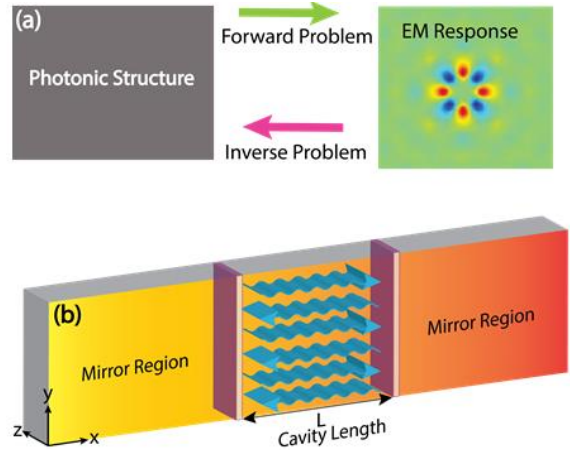

Fig. 1 (a) Schematic representation of forward and inverse problem approach for photonic structure design and (b) parallel reflector resonator.

In our resonator design work, during the inverse optimization process the plane reflector conditions are defined to create symmetric reflection effect where waves are intended to be reflected repeatedly to create the resonance at the targeted wavelength. The length of the cavity region is chosen to support the fundamental transverse electric (TE) mode in accordance with the number of supported modes that are stated in Eq. 5.

III. RESONATOR DESIGN AND NUMERICAL RESULTS

In this study, the design process of the resonating structure starts with a bulky waveguide slab structure shown in Fig. 2(a). In the OF algorithm, the wave source with the Gaussian intensity profile located at the center of the bulky waveguide structure is defined as the input source $\vec{H}_{in}$ and the light wave propagates in both –x and +x directions. Also, boundary conditions are defined to reflect the light at the wavelength of 1550 nm to create a cavity region. Finally, we set the initial condition of the optimization routine to a uniform permittivity level $\varepsilon_{init}$ as 12.25. This parameter has a considerable influence on the optimized design structure and performance. The objective of the design is defined to create reflective medium (Bragg-reflection like region) embedded in the mirrors.

The inverse optimization is applied in a way of alternating directions of the linear sub-problems, which are dependent on H-field and $\varepsilon$, separately, and extracted from the electromagnetic master equation [21]. The ultimate resonator design having continuous dielectric distribution between values of 1.0 and 12.25 is demonstrated in Fig. 2(b). The obtained resonator consists of almost symmetrically distributed mirror segments as can be seen in Fig. 2(b). Here, the air-filled sections gradually increase in size with the distance from the center that is intentionally revealed by the algorithm. After obtaining resonating structure we performed binarization process to generate realistic photonic device as shown in Fig. 2(c).

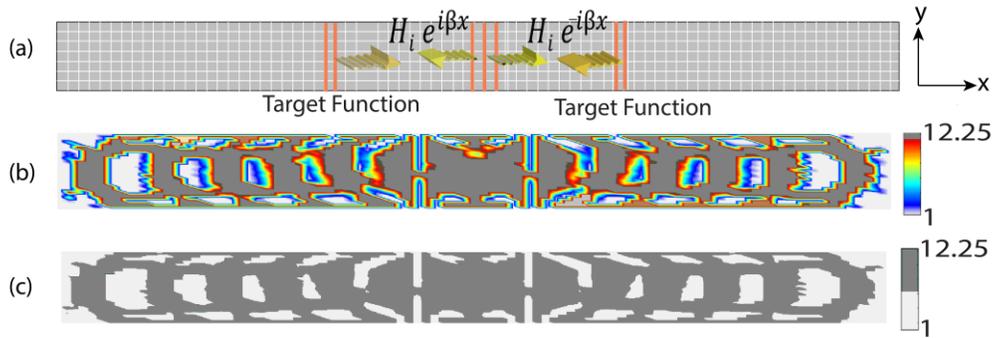

Fig. 2. (a) Schematic representation of the defined waveguide medium and target functions. (b) Designed continuous index resonator structure, and (c) designed binary resonator structure.

As seen in Fig. 3(a), we investigated the changes of physical residual respect to the iteration number to observe the success of the design. The footprint of the structure is chosen as 13.394 x 0.592 µm². Physical residual converges vary fast up to 600 iterations. Here we discuss 100 iterations and 800 iterations design result which are discretized. Design results are investigated via the FDTD method[28]. As seen in Fig. 3(b) the algorithm starts to create Bragg-like layers. In addition, as represented in Fig. 3(c) the desired mode cannot be confined at cavity region yet. On the other side, as can be seen in Fig. 3(d), there are more two layers at the structure result for 800 iterations and the algorithm can create a bigger region. As can be seen from the Fig.3.e), fundamental TE mode begin to confine at cavity region to perform the design function.

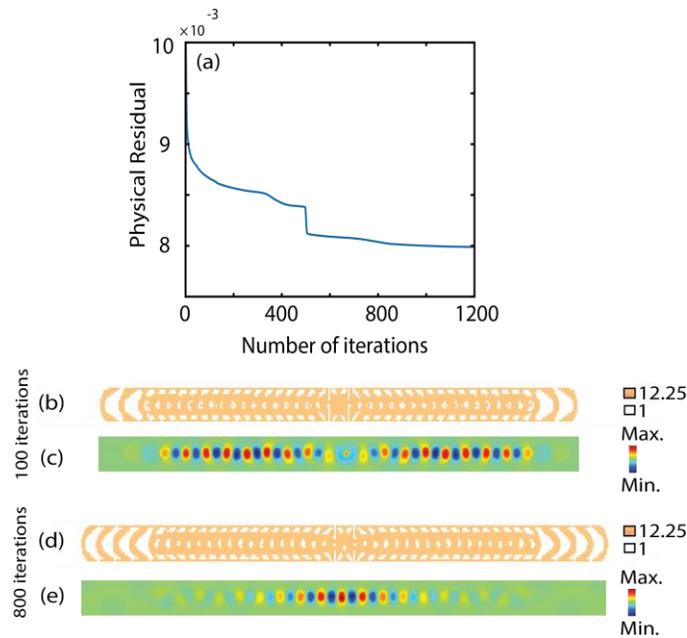

Fig. 3. (a) Physical residual versus number of iterations, (b) The designed structure and the corresponding H-field intensity (c) at 1550 nm for 100 iterations, (d) the designed structure and the corresponding H-field intensity (e) at 1550 nm for 800 iterations.

In this study, we discuss the result of 1200 iterations in detail and the design with the appropriate binarization condition previously described in the associated section is shown in Fig. 4(b). The results of the design are analysed by FDTD method and wideband transmission spectrum is shown in Fig. 4(a). As a natural result of the parallel plane mirror resonator, discrete set of the resonator modes are formed. Fundamental TE mode ($M_1$) is confined at the cavity region which is shown in Fig. 4(c) at the desired wavelength. We also investigate the $H$-field intensity of the transverse resonator modes ($M_{2-6}$) and show that they are not localized at the cavity region which can be seen from Figs. 4(c)-(h). The quality factor value of the cavity is calculated as around $\sim 24 \times 10^3$ at the resonance wavelength, 1558.85 nm.

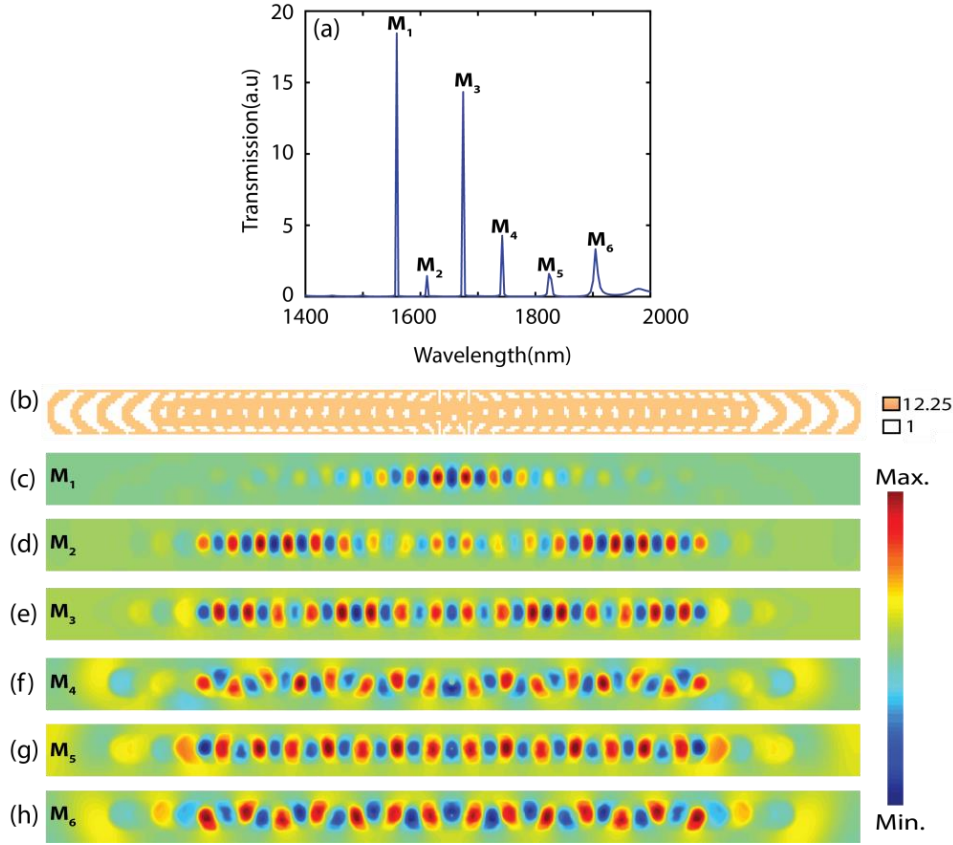

Fig. 4. (a) Transmission spectrum of the final design (b) Designed structure with a foot-print of 13.394 x 0.592 μm² and the corresponding $H$-field distribution of (c) $M_1$ at λ=1558.85 nm, (d) $M_2$ at λ=1611.50 nm (e) $M_3$ at λ=1674.33 nm (f) $M_4$ at λ=1742.226 nm (g) $M_5$ at λ=1829.65 nm. (h) $M_6$ at λ=1904,53 nm.

## IV. CONCLUSION

In conclusion, parallel mirror resonator design was proposed and demonstrated for the first time by using a recently-emerged highly promising objective-first inverse design method. In this respect, parallel dielectric mirror-based resonator design confining the light at the predefined resonance wavelength of 1550 nm has been achieved. The structure is manufacturable through the standard lithography techniques. The modified algorithm determines the size and shape of the mirror sections in the design to obtain the optimal Bragg-like reflection condition. Therefore, the design exhibiting the expected field distribution is accomplished without any traditional parametric analysis for the features of the mirror-like sections. The parameters defined in the cost function, and the dielectric distribution ($\varepsilon$) of the structure is obtained in an iterative manner. The wave equation presents a non-convex scalability of the design dimensions and in-advance determination of the resonance wavelength in the algorithm are other significant characteristics of the utilized method. Parallel mirror resonator design cavity structures presented in this study have great potential to facilitate nanoscale optical filters that can be utilized for implementing light sources and biosensors.

ACKNOWLEDGEMENT

The authors acknowledge the financial support of the Scientific and Technological Research Council of Turkey (TUBITAK) (116F200) and the partial support of the Turkish Academy of Sciences.